\documentstyle[aps,prl,subfigure,epsfig,amsmath,amssymb]{revtex}

\begin{document}
\draft

\title{Exact particle and kinetic energy densities for one-dimensional 
  confined gases of non-interacting fermions}

\author{ Patrizia Vignolo, Anna Minguzzi and M. P. Tosi}
\address{  Istituto Nazionale di Fisica della Materia and Classe di Scienze,
Scuola Normale Superiore,
Piazza dei Cavalieri 7, I-56126 Pisa, Italy}
\maketitle
\begin{abstract}
  We propose a new method for the evaluation of the particle density and 
  kinetic pressure profiles in inhomogeneous one-dimensional systems
  of non-interacting fermions, and apply it to harmonically confined systems of
  up to $N=1000$ fermions. The method invokes a Green's function 
  operator in coordinate
  space, which is handled by techniques originally developed for the
  calculation of the density of single-particle states from Green's functions
  in the energy domain. In contrast to the Thomas-Fermi (local density) 
  approximation, the exact profiles under harmonic confinement 
  show negative local pressure 
  in the tails and a prominent shell structure which may become 
  accessible to  
  observation in magnetically trapped gases of fermionic alkali atoms.
\end{abstract}
The 
techniques which have led to the achievement of Bose-Einstein
condensation in 
vapors of bosonic atoms \cite{bec_rb,bec_na,bec_li,bec_h} are 
currently being used to trap and
cool dilute gases  of fermionic alkali atoms \cite{fermioni}.
Under magnetic confinement the $s$-wave collisions between pairs
of fermions in a single hyperfine level are suppressed by the
Pauli principle, and the geometry of the trap can be adapted to 
have cylindrical symmetry with a transverse confinement which may be 
hundreds of times stronger than the longitudinal one.
It is thus possible to experimentally realize quasi-onedimensional
inhomogeneous systems of almost non-interacting fermions at very
low temperature and high purity.

A number of one-dimensional (1D) physical  models 
can be solved exactly \cite{lieb}
and their solution 
serves as a test of approximate theories and contributes to the
understanding of real systems.
Some important examples are the determination of  the ground state and
excitation spectrum of a hard-core Bose gas in 1D \cite{lieb_liniger}
and the solution of the 
Kronig-Penney model for the electron energy bands
in a 1D crystal lattice \cite{kronig}. 
The spectral and transport properties of 
other 1D systems of non-interacting electrons have
been studied as models for polymers and quantum
wires, 
using Green's function methods
\cite{haydock,economu,pino,qw} 
for which ingenious techniques such as a decimation/renormalization
procedure \cite{rino1,rino2} have been developed.

In this work we present a new method for the exact evaluation of the 
ground-state particle density profile in a spin-polarized
1D system of up to large numbers of non-interacting
fermions in arbitrary spatial confinement. In essence we show that, just
as the single-particle density of states in the energy domain can be
obtained by powerful Green's function methods, similar 
techniques yield
the particle density in the 1D space domain. 
In fact, our method also allows the evaluation
of higher moments of the one-body density matrix: we focus here on its second
moment, which is simply
proportional to the kinetic energy density and to the kinetic pressure.

As an application of the general method we give results for the
particle density and
kinetic pressure profiles of a degenerate Fermi gas in harmonic confinement. 
This model is directly relevant to the current experiments on 
atomic Fermi gases and we show that the shell structure noticed for the
particle density in earlier theoretical studies in 3D 
\cite{schneider,bruun} is greatly enhanced in 1D.
We also use our exact results to test the Thomas-Fermi (local density) 
approximation
in dependence of the number of fermions in the confined gas.

\section{General formulation}
The one-body Dirac density matrix for a system of $N$ non-interacting fermions
at zero temperature can be expanded on the single-particle 
wavefunctions
$\psi_i(x)=\langle x\,|\psi_i\rangle$ 
as $\rho(x_1,x_2)=\sum_{i=1}^N \psi_i^*(x_1)
\psi_i(x_2)$. By using the representation of the translation operator
this becomes
\begin{equation}
\rho(x_1,x_2)=\sum_{i=1}^N \psi_i^*(x_1)e^{-i \hat p(x_1-x_2)}
\psi_i(x_1)\;,
\end{equation}
showing how distant points are
correlated through the momentum operator $\hat p$. Expansion
in powers of
the relative coordinate $r=x_1-x_2$ yields physical
observables such as the particle density profile $n(x)$,
\begin{equation}
n(x)=\left. \rho(x+r/2,x-r/2)\right|_{r=0}=
\sum_{i=1}^N \langle \psi_i\,|\,\delta(x-x_i)\,|\,\psi_i\rangle\;
\label{ndef}
\end{equation}
and the kinetic pressure $P(x)$, 
\begin{equation}
P(x)=-\frac{\hbar^2}{m} \frac{\partial^2}{\partial r^2} \left.
\rho(x+r/2,x-r/2) \right|_{r=0} = \frac{1}{2m} \sum_{i=1}^N 
\langle \psi_i|
p_i^2\delta(x-x_i)+\delta(x-x_i) p_i^2|\psi_i \rangle\;. 
\label{pdef}
\end{equation}
This is twice the kinetic energy density.

The main idea of this Letter is to rewrite Eqs. (\ref{ndef}) and
(\ref{pdef}) as the imaginary part of the ground-state average
of suitable operators
related to the Green's
function in 
coordinate space $G(x)=(x-\hat x + i\varepsilon )^{-1}$.
We have  
\begin{equation}
n(x)=-\frac{1}{\pi}\lim_{\varepsilon\rightarrow0^+} {\rm
Im}\,\sum_{i=1}^N \langle \psi_i\,|G(x)|\,\psi_i\rangle\;
\label{ndefnoi}
\end{equation}
and
\begin{equation}
P(x)=-\frac{1}{\pi}\lim_{\varepsilon\rightarrow0^+} {\rm
Im}\,\sum_{i=1}^N \langle \psi_i\,|\hat \frac{p^2}{m} G(x)|\,\psi_i\rangle\;.
\label{pdefnoi}
\end{equation}
$G(x)$ can then be treated by methods analogous to those
used for
treating Green's functions in the energy domain. 

The equivalence between expressions (\ref{ndef}) and (\ref{ndefnoi})
is easily proved in the
coordinate representation, where the density profile in Eq. (\ref{ndefnoi})
reads
\begin{equation}
n(x)=-\frac{1}{\pi}\lim_{\varepsilon\rightarrow0^+} {\rm
Im\,}\sum_{i=1}^N \int {\rm d}x_i
|\psi_i(x_i)|^2\frac{1}{x-x_i+i \varepsilon}\;,
\end{equation}
yielding Eq.~(\ref{ndef})
when one takes the limit $\varepsilon\rightarrow0^+$.
The equivalence between expressions (\ref{pdef}) and (\ref{pdefnoi})
for $P(x)$ is similarly proved.

Evidently, this method can be applied to all 1D systems which may be described 
by single-particle
orbitals: one only needs to know the representation of the position 
and momentum 
operators on such a basis. 
Hence, interactions could also be included
in evaluating the particle density through the use of Kohn-Sham 
single-particle
orbitals. Models of displacement fields (such as those induced by
impurities) may also be studied directly without previous evaluation
of orbitals.

\section{Non-interacting Fermi gas in harmonic trap}
As already noted, a 1D Fermi model is relevant to
the spin-polarized fermionic
vapors in magnetic confinement \cite{fermioni}, where
it is possible to realize
experimentally a 1D configuration by making use of very
anisotropic axially symmetric traps. 
At low
temperature only the transverse ground state of the trap is populated
and the vapor can be described by an effective 1D
harmonic Hamiltonian. 

An analytic expression for the particle density 
of this system
has been given by Husimi \cite{husimi,march} in terms of the
wavefunction of the $N$-th fermion in the trap. However, a
calculation of the density profile and the kinetic pressure 
by his approach is limited to small values of $N$ and has been reported
only for $N=$ 1 and 2 \cite{march}. Our method allows us to
efficiently evaluate these ground state properties even for
quite large numbers of particles.
Of course, the simplicity of the representation of the position
and momentum operators in this system makes it a favorable example.
\subsection{Particle density profile} 
We start with the calculation of the particle density.
For a linear harmonic oscillator the position operator is represented
as 
$\hat{x}=(a+a^{\dag})/\sqrt 2$
on the basis $\{|\,\psi_i\rangle\}$ of the
eigenvectors of the Hamiltonian. As usual, the creation and
destruction operators satisfy the relations
$a\,|\,\psi_i\rangle=\sqrt{i-1}\,|\,\psi_{i-1}\rangle$
and
$
a^{\dag}\,|\,\psi_i\rangle=\sqrt{i}\,|\,\psi_{i+1}\rangle$.
A straightforward procedure for evaluating the profile
(\ref{ndefnoi}) is to invert
 the matrix $(x-\hat{x}+i\varepsilon)$
and to calculate its trace on the submatrix of its first
$N\times N$ block (Tr$_N$). 
We employ  the relation
${\rm Tr}_NQ=\partial
\left[\ln\det(Q^{-1}+\lambda{\Bbb{I}}_N)\right]/\partial\lambda|_{\lambda=0}$,
where ${\Bbb{I}}_N$ is a diagonal semi-infinite matrix with
its first $N$ eigenvalues equal to 1 and null elsewhere \cite{formula},
to obtain 
\begin{equation}
n(x)=-\frac{1}{\pi}\lim_{\varepsilon\rightarrow0^+}{\rm Im}
\frac{\partial}{\partial\lambda}
\left[\ln\det(x-\hat{x}+i\varepsilon+\lambda{\Bbb{I}}_N)\right]_{\lambda=0}.
\label{dens}
\end{equation}
The calculation of the determinant in Eq. (\ref{dens}) is conveniently
performed by the recursive algorithm developed in
\cite{fgv}. Renormalization of the tridiagonal
operator $\hat{R}=\hat{x}-\lambda{\Bbb{I}}_N$  allows us to 
write  
\begin{equation}
\det{(x-\hat{R}+i\varepsilon)}=\prod_{k=1}^{\infty}
(x-\tilde{a}_k+i\varepsilon)
\label{prod1}
\end{equation}
with $\tilde{a}_1=-\lambda$,
$\tilde{a}_{k+1}=-\lambda+\frac{1}{2}k(x-\tilde{a}_{k}+i\varepsilon)^{-1}$ 
for $1<k<N$ and
$\tilde{a}_{k+1}=\frac{1}{2}k(x-\tilde{a}_{k}+i\varepsilon)^{-1}$ 
for $k\geq N$.

The scheme given in Eqs. (\ref{dens}) and (\ref{prod1}) is easily implemented 
numerically. In practice we have performed the calculation of the determinant
(\ref{prod1}) up to the product of its first $M$ terms,
which corresponds to inverting an $M\times M$ matrix.
We have checked the convergence of this approximation by increasing
the dimension $M$ and correspondingly decreasing the value of
$\varepsilon$.

\begin{figure}
\epsfig{file=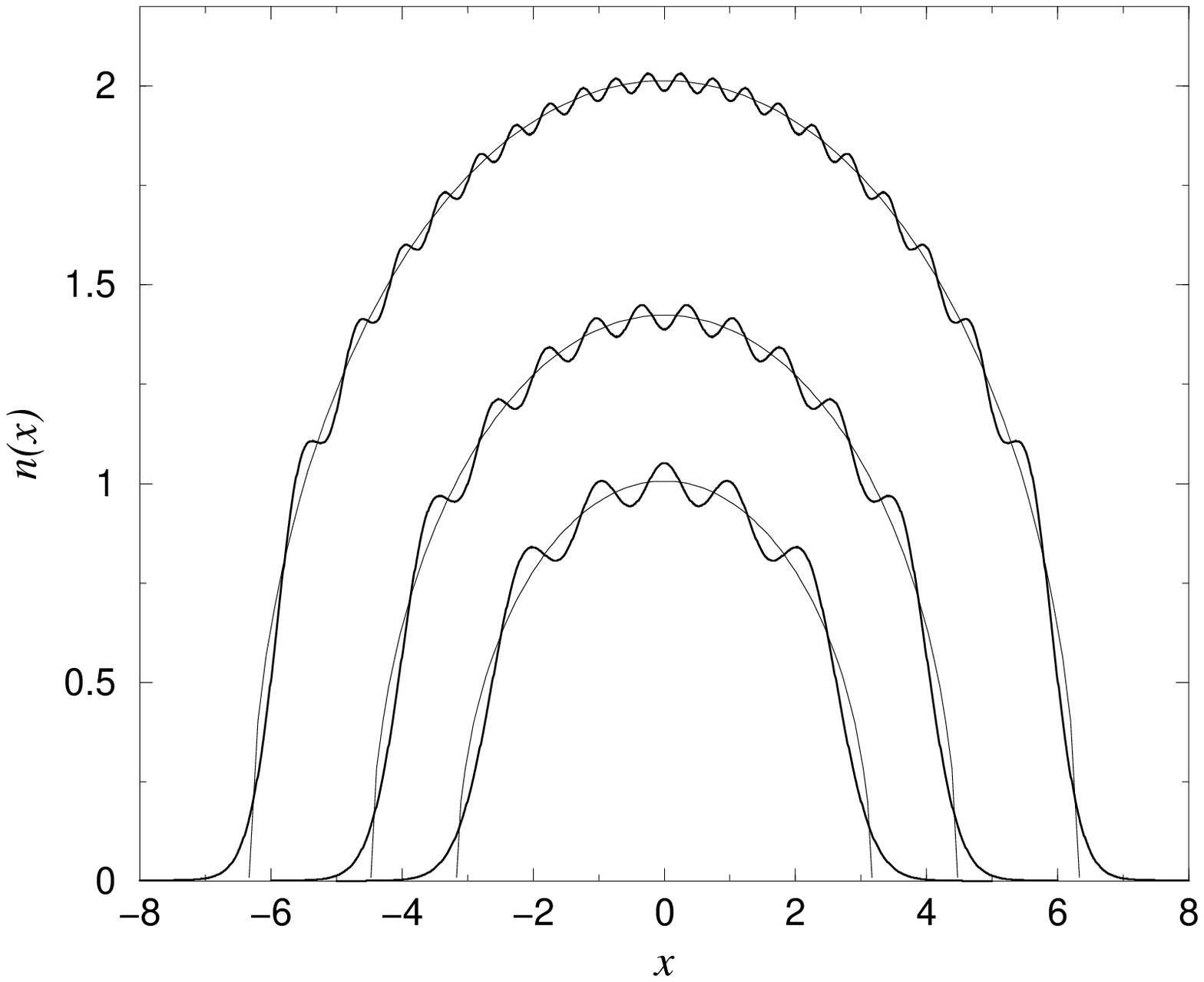,width=1\linewidth}
\caption{Exact particle density profile (bold lines) for $N=5$,
10 and 20 harmonically confined fermions, 
compared with the corresponding profiles
evaluated in the local density approximation.
Positions are in units of the characteristic length
of the harmonic oscillator $a_{ho}=\sqrt{\hbar/(m \omega)}$ and
the particle density in units of $a_{ho}^{-1}$.}
\label{fig1}
\end{figure}

In Fig.~\ref{fig1} we report the density profile $n(x)$ for $N=5$,
10 and 20 fermions,
with $M=10^5$ and $\varepsilon=0.01$. 
The exact profiles are also compared with those given by the 
Thomas-Fermi approximation (LDA),
\begin{equation}
n_{\rm LDA}(x)=\frac{1}{\pi}(2N-x^2)^{1/2}
\end{equation}
(in units such that $\hbar=1$, $m=1$ and
$\omega=1$).
The exact profiles contain $N$ oscillations, which
become smaller in relative amplitude as $N$ increases. 
Without any special numerical efforts we have evaluated the exact
density profile up to $N=1000$: this would otherwise require the calculation
of Hermite polynomials up to the 1000$^{th}$ degree.
For large $N$ the oscillations are so small
in relative amplitude that their smoothing in
the LDA profile becomes reasonably accurate,
except for the region of the tails.

\subsection{Kinetic pressure profile}
The particle density profile that we have evaluated above 
is the analogue of the density of single-particle 
states in the energy domain. 
In the space domain other single-particle quantities acquire physical
interest, as is the case for the kinetic pressure in
Eq.~(\ref{pdef}). 
We show how also for this function one can profitably
resort to a renormalization
technique.

Taking $Q={\hat{p}}^2G(x)$, Eq.~(\ref{pdefnoi}) can be written as
\begin{equation}
P(x)=-\frac{1}{\pi}\lim_{\varepsilon\rightarrow0^+}{\rm Im}
\frac{\partial}{\partial\lambda}
\left[\ln\det(x-\hat{x}+i\varepsilon+\lambda{\Bbb{I}}_N{\hat{p}}^2)
\right]_{\lambda=0}.
\label{penta}
\end{equation}
The matrix $(x-\hat{x}+i\varepsilon+\lambda{\Bbb{I}}_N{\hat{p}}^2)$ 
appearing  in Eq. (\ref{penta})
is pentadiagonal on the first $N$ rows, 
owing to the form of the operators $\hat{x}$ and
$\hat{p}=i(a^{\dag}-a)/\sqrt 2$ in the basis of the energy eigenstates
$\{|\,\psi_i\rangle\}$.
The calculation of the determinant of such a matrix 
can again be performed by the recursive algorithm given in \cite{fgv}. 
Renormalization of the operator $\hat{K}=
\hat{x}-\lambda{\Bbb{I}}_N{\hat{p}}^2$ is made on blocks of dimension 2
for the pentadiagonal part and on blocks of dimension 1 for
the tridiagonal part. This
allows us to 
write 
\begin{equation}
\det(x-\hat{K}+i\varepsilon)=\left\{\begin{array}{l}
\prod_{j=1}^{(N+2)/2}\det(x-\tilde{A}_j+i\varepsilon)
\prod_{k=N+3}^{\infty}(x-\tilde{a}_k+i\varepsilon)
\qquad ({\rm even}\; N)\\[3mm]
\prod_{j=1}^{(N+3)/2}\det(x-\tilde{A}_j+i\varepsilon)
\prod_{k=N+4}^{\infty}(x-\tilde{a}_k+i\varepsilon)
\qquad 
({\rm odd}\; N)\;.\\
\end{array}\right.
\label{ricorr}
\end{equation}
The renormalized $2\times 2$ 
blocks $\tilde{A_j}$ satisfy the recursion relation 
\begin{equation}
\tilde{A_j}=A_j+B_{j,j-1}(x-\tilde{A}_{j-1}+i\varepsilon)^{-1}B_{j-1,j}
\end{equation}
for $j>1$ and $\tilde{A}_1=A_1$.  The matrices $A_j$,
$B_{j,j+1}$ and $B_{j+1,j}$ are 
submatrices of the operator $\hat{K}$, which are defined as follows:
\begin{equation}
A_j=\left(\begin{array}{cc}
-\lambda\left(2j-{3}/{2}\right)\theta_{N-2j+1}&
\sqrt{(2j-1)/{2}}\\
\sqrt{(2j-1)/{2}}&
-\lambda\left(2j-1/2\right)\theta_{N-2j}\\
\end{array}
\right),
\end{equation}
\begin{equation}
B_{j,j+1}=\left(\begin{array}{cc}
\lambda\sqrt{2j(2j-1)}/2\,\theta_{N-2j+1}&0\\
\sqrt{j}&\lambda\sqrt{2j(2j+1)}/2\,\theta_{N-2j}\\
\end{array}
\right)
\end{equation}
and
\begin{equation}
B_{j+1,j}=\left(\begin{array}{cc}
\lambda\sqrt{2j(2j-1)}/2\,\theta_{N-2j-1}&\sqrt{j}\\
0&\lambda\sqrt{2j(2j+1)}/2\,\theta_{N-2j-2}\\
\end{array}
\right),
\end{equation}
with $\theta_k=1$ for $k\geq 0$ and $\theta_k=0$ otherwise.
The recursion relation for the elements $\tilde{a}_k$
is again $\tilde{a}_{k+1}=\frac{1}{2}k/(x-\tilde{a}_k+i\varepsilon)$,
the first elements being
$\tilde{a}_{k+1}=\frac{1}{2}k\{x+i\varepsilon-[\tilde A_{k/2}]_{22}-
[\tilde A_{k/2}]_{21}\cdot [\tilde A_{k/2}]_{12}
/(x+i\varepsilon-[\tilde A_{k/2}]_{11})\}^{-1}$ with
$k=N+2$ for even $N$ and $k=N+3$ for odd $N$.
We have studied the convergence of
the determinant in Eq. (\ref{ricorr}) as for the case
of the particle density profile.

\begin{figure}
\epsfig{file=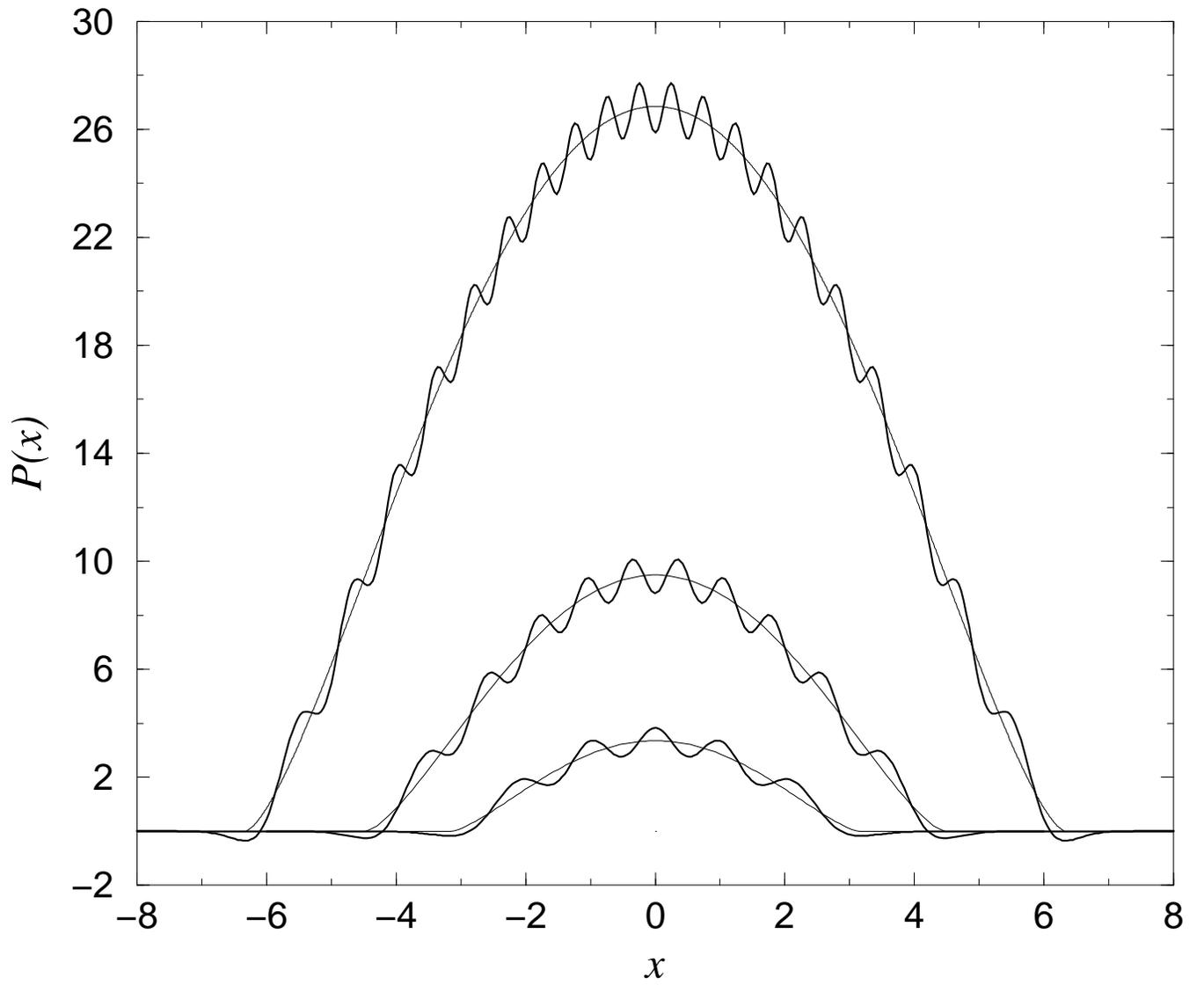,width=1\linewidth}
\caption{Exact kinetic pressure  profile (bold lines) for $N=5$,
10 and 20 harmonically confined fermions, 
compared with the profiles
evaluated in the local density approximation.
Positions are in units of $a_{ho}=\sqrt{\hbar/(m \omega)}$ and
the kinetic pressure in units of $\hbar \omega a_{ho}^{-1}$.}
\label{fig2}
\end{figure}

\begin{figure}
\epsfig{file=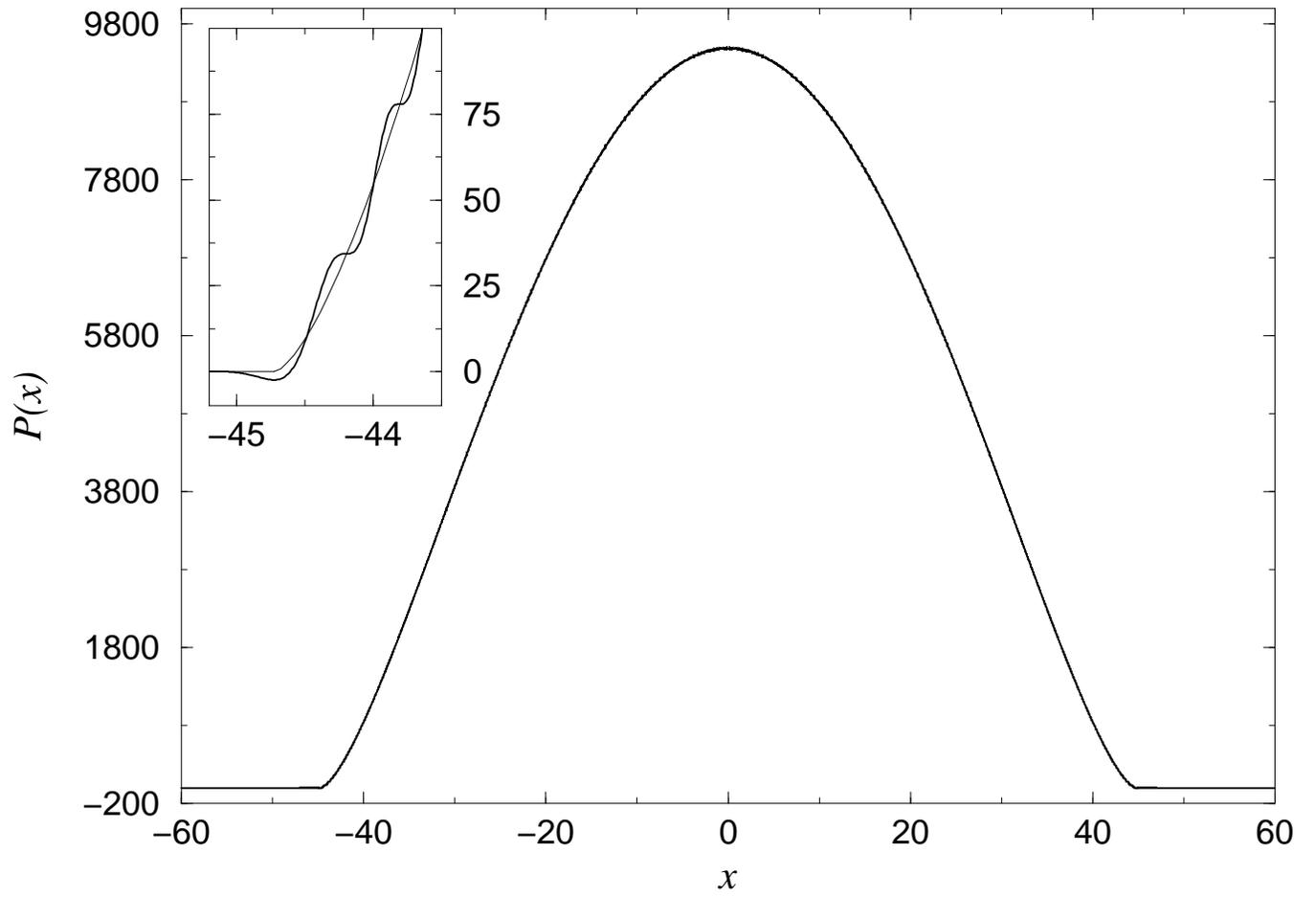,width=1\linewidth}
\caption{Exact kinetic pressure  profile (bold lines) for $N=1000$,
compared with the profile
evaluated in the local density approximation.
The inset shows an enlarged view of the tail of the profiles.
The units are as in Figure \ref{fig2}. In this calculation we have
employed a matrix of dimension $M=10^7$ and 
chosen $\varepsilon=10^{-3}$ (see notations in the text).}
\label{fig3}
\end{figure}

In Fig.~\ref{fig2} the kinetic pressure $P(x)$ is plotted for $N=5$,
10 and 20 with $M=10^5$ and $\varepsilon=0.01$, together with
the profiles $P_{\rm LDA}(x)$ evaluated in
the Thomas-Fermi approximation,
\begin{equation}
P_{\rm LDA}=\frac{1}{3\pi}(2N-x^2)^{3/2}.
\end{equation}
The exact kinetic pressure 
shows $N$ oscillations
and has the peculiarity of being negative in the tails.
This microscopic quantum effect, which is missing in the local density
description, reflects the fact that in the  low density region the
kinetic energy  decreases with increasing density.
We have checked that our results agree with those reported in
\cite{march} for $N=1$ and 2,
and carried out the calculation of $P(x)$ up to $N=1000$.
The kinetic pressure profile for N=1000, as shown in Fig.~\ref{fig3},
looks almost 
indistinguishable from the LDA prediction but stills presents a region
of negative kinetic pressure in the tails (inset in Fig.~\ref{fig3}).

In conclusion, in this Letter we have given a general formula for the exact
particle density and kinetic pressure profiles of 
a 1D many-fermions system in terms of a Green's
operator in coordinate space. We have made use of the
decimation/renormalization procedure and of other recursive techniques,
originally developed to evaluate the spectral properties of
quasi-1D systems in solid state physics, to efficiently
calculate the exact density profiles of a harmonically confined
non-interacting Fermi gas.
Within the same general scheme the particle density could also
be evaluated by employing a suitable Kirkman-Pendry 
relation \cite{pendry}, as will be reported elsewhere.
We have verified that for large number of atoms ($N=1000$) 
the local density approximation reproduces reasonably well
the exact profiles except for the region of the tails, where the
exact kinetic pressure is negative. 

We believe that the present method opens the way for a novel approach
to the
equilibrium properties of spatially
inhomogeneous 1D systems. The expressions here derived can be extended
to finite temperature and to calculate partial density profiles
for subgroups of atoms. The kinetic energy 
density functional can be studied through the calculation of the function 
$P[x(n)]/2$ where $x(n)$ is obtained by local inversion of the exact 
profile $n(x)$. Pressure fluctuations will become accessible
to study through the 
evaluation of higher moments of the one-body density matrix.
The density profiles of the harmonically trapped Fermi gas in 1D, 
showing a prominent shell
structure as displayed in our calculations, could become observable in 
experiments on alkali vapours.

\acknowledgements
It is a pleasure to thank Professor N. H. March for useful discussions on
Fermi vapors and for drawing our attention to the work published in
Refs. \cite{husimi,march}. One of us
(P. V.) would like to thank Professor G. Grosso and Dr R. Farchioni
for fruitful discussions.

\newpage

\newpage

\end{document}